\documentclass[useAMS,usenatbib]{mn2e}

\usepackage{graphicx}% Include figure files
\usepackage{dcolumn}% Align table columns on decimal point
\usepackage{bm}% bold math
\usepackage{color}

\usepackage{verbatim}
\begin{document}
% \title{Cosmic Rays from Magnetars}
% \author{Jeremy S. Heyl}
% \thanks{Canada Research Chair}
% \email{heyl@phas.ubc.ca}
% \author{Ramandeep Gill}
% \affiliation{Department of Physics and Astronomy, 6224 Agricultural
%  Road, University of British Columbia, Vancouver, British Columbia,
%  V6T 1Z1}
% \author{Lars Hernquist}
% \affiliation{Harvard-Smithsonian Center for Astrophysics, 60
%  Garden St., Cambridge MA 02138}
% \date{\today}

\title{Cosmic Rays from Pulsars and Magnetars}
\date{Accepted ---.  Received ---; in original form ---}
\author[J. S. Heyl, R. Gill and L. Hernquist]{Jeremy
  S. Heyl$^{1\dagger}$, Ramandeep Gill$^1$ and 
Lars Hernquist$^2$\\
$^{1}$Department of Physics and Astronomy, University of British
Columbia, Vancouver, British Columbia, Canada, V6T 1Z1 \\
$^{2}$Harvard-Smithsonian Center for Astrophysics, 60
  Garden St., Cambridge MA 02138\\
$^{\dagger}$Email: heyl@phas.ubc.ca; Canada Research Chair
}
% \thanks{Canada Research Chair}
% \email{heyl@phas.ubc.ca}
% \author{Ramandeep Gill}
% \affiliation{Department of Physics and Astronomy, 6224 Agricultural
%  Road, University of British Columbia, Vancouver, British Columbia,
%  V6T 1Z1}
% \author{Lars Hernquist}
% \affiliation{Harvard-Smithsonian Center for Astrophysics, 60
%  Garden St., Cambridge MA 02138}
% \date{\today}

\pagerange{\pageref{firstpage}--\pageref{lastpage}} \pubyear{2009}

\maketitle

\label{firstpage}
\begin{abstract}
  We compare the expected abundance of cosmic-ray electrons and
  positrons from pulsars and magnetars.  We assume that the
  distribution of infant pulsars and magnetars follows that of
  high-mass stars in the Milky Way and that the production rate of
  cosmic rays is proportional to the spin-down and magnetic-decay
  power of pulsars and magnetars, respectively.  In combination with
  primary and secondary cosmic-ray leptons from other sources
  (especially supernova remnants), we find that both
  magnetars and pulsars can easily account for the observed cosmic-ray
  spectrum, in particular the dip seen by HESS at several TeV and the
  increase in positron fraction found by PAMELA.
\end{abstract}

\section{\label{sec:intro}Introduction}

%\homepage[]{Your web page}
%\altaffiliation{}

Recent measurements of cosmic rays from energies of one GeV to one TeV
have shown a spectrum \citep{2008Natur.456..362C,2008PhRvL.101z1104A}
and abundance of positrons \citep{2009Natur.458..607A} that are
difficult to reconcile with astrophysical sources.  These observations
have naturally stirred up great interest among the particle physics
community and especially people who study dark matter
\citep[e.g.][]{2009NatPh...5..176H}. Essentially the problem is that
the observed spectrum is much harder than that expected to be produced
by astrophysical sources such as supernova shocks or pulsars, so it is
quite natural to look at more exotic possibilities, including decaying
dark matter. On the other hand, several works have also examined the
question of whether or not astrophysical sources can account for the
observed spectrum and positron abundance by including a reasonable
model for the diffusion of cosmic rays through the Galaxy. In this
Letter we consider the expected abundance of cosmic-ray electrons and
positrons (hereafter referred to as cosmic rays) produced by
magnetars and pulsars.

Magnetars \citep{Dunc92,Thom95} are ultramagnetized neutron stars with
fields $B_* \sim 10^{14} - 10^{16}$ G, fueled by either magnetic field
decay \citep{Thom96,Heyl98decay} or residual thermal
energy \citep{Heyl97kes,Heyl97magnetar}.  Magnetars are thought to
produce copious numbers of electron-positron pairs both in
bursts \citep{Thom95,Heyl03sgr} and
in quiescence \citep{Heyl05sgr,2007ApJ...657..967B}.  
Moreover, the local radiation
from these pairs is thought to power the bursts from soft-gamma
repeaters \citep{Thom95} as well as the recently discovered non-thermal
emission from soft-gamma repeaters and anomalous x-ray pulsars in
quiescence \citep{2005A&A...433L...9M,2006ApJ...645..556K}.
Consequently, magnetars seem a natural candidate to make an important
contribution to the observed abundance of electrons and positrons
observed at Earth.

\citet{2008arXiv0812.4457P} presented a detailed calculation of
the expected spectrum of cosmic rays in various models for their
production from pulsars. His calculation
focused on the population of gamma-ray pulsars. The situation for
magnetars is somewhat different because these objects are much rarer
than gamma-ray pulsars; the birth rate for magnetars is roughly ten
times smaller than that of pulsars.  Consequently, the cosmic ray flux
that we observe can come both from the magnetars that are active today
as well as from more ordinary looking objects that may have been
active magnetars in the past such as
RXJ~0720.4-3125 \citep{Heyl98rxj,Heyl98decay}.  To calculate the
expected abundances of cosmic rays from magnetars, we generate a Monte
Carlo simulation of the magnetar population over the past ten million
years and compare it with a similar simulation of the pulsar
population over the past million years.

\section{The Models}

\subsection{Emission}

To model the cosmic ray emission from pulsars we assume a production
rate always proportional to the spin-down power by magnetic dipole
emission.  In contrast to \citet{2008arXiv0812.4457P}, we do not make
any distinction between objects that we expect to be gamma-ray pulsars
versus the rest of the pulsar population.  To specify the model, we
assume that all the pulsars are born with a period of 49 milliseconds
and spin down to one second by the time we observe them. For the
magnetars, we take the cosmic ray emission to be proportional to the
magnetic field decay power. We use the calculations of
\citet{Heyl98decay} to follow the evolution of the magnetic field
starting at $10^{16}$~G at the pole on the surface.  We assume that
only the energy in the external magnetic field is converted into
cosmic rays and that the external field is strictly dipolar (if the
internal field is uniform, there is a factor of fifty more energy
within the star).  For ease of comparison, the initial spin period of
the pulsar population is chosen such that the total spin-down energy
in the pulsars equals the total magnetic field energy of the
magnetars.\footnote{We note that these choices, while somewhat
  arbitrary, are motivated by observed characteristic properties of
  young and old pulsars and magnetars.  Moreover, variations in pulsar
  spin periods and the magnetic energy available to produce cosmic
  rays from magnetars affect only the normalization of the cosmic ray
  flux seen at Earth, not their observed energy spectrum, which is
  mostly sensitive to the distribution of sources in space and with
  time and the spectral index of the intrinsic emission process.}  The
magnetars and pulsars may inject high-energy electrons and positrons
into interstellar space.  We assume that the fraction of the magnetic
field or spin-down energy converted to pairs is given by $\epsilon_m$
and $\epsilon_p$ respectively.  These are determined by insisting that
total electron-positron cosmic-ray abundance from other expected
sources \citep{2010arXiv1002.1910D} when combined with either the
magnetars or pulsars does not exceed that observed by Fermi around
300~GeV.

\begin{figure}
\includegraphics[width=8.5cm,clip]{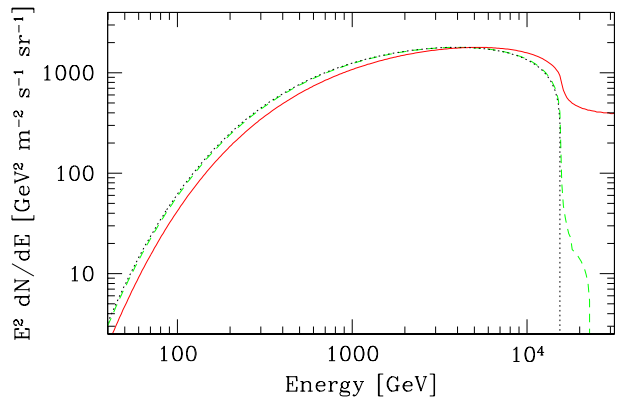}
\caption{The spectrum of cosmic rays from a single pulsar (green
 dashed curve), magnetar (red solid curve) or instantaneous burst
 (black dotted curve) at a distance of 200~pc and an age of 20,000
 years under the ``LOW'' scenario. The total energy release is the
 same for the three models.}
\label{fig:compare}
\end{figure}

An important difference between magnetars and pulsars is the evolution
of cosmic ray generation with time. In the case of a pulsar the bulk
of the cosmic rays are made during the very early life of the pulsar
as it is spinning down most rapidly. On the other hand, we assume that
magnetars generate cosmic rays from magnetic field decay, which is
also thought to power thermal and nonthermal photon emission.  The
magnetic field decay timescale is typically several thousand years so
the emission of cosmic rays from the magnetar is of longer
duration. This difference shows up in the observed spectrum of cosmic
rays from a nearby magnetar as compared to a nearby young pulsar
(Fig.~\ref{fig:compare}). In the case of a pulsar, one sees a
relatively smooth spectrum up to the energy of the electrons or
positrons that have had time to cool since the birth of the pulsar; at
this energy there is a relatively sharp cut off. In the case of the
\citet{2008arXiv0812.4457P} calculation this cut off is infinitely
sharp. A magnetar releases its magnetic energy gradually, so the
cosmic rays that we observe at Earth have a variety of ages, and the
energy cut off is much less sharp.  However, this is only a
second-order effect in determining the shape of the spectrum from an
ensemble of objects; the largest differences in the expected spectrum
from a collection of magnetars versus pulsars owes to the relative
ages of these two populations.

A second important issue is the spectrum of the cosmic rays where they
are produced. In both cases we assume that the number of cosmic ray
particles decreases as $\gamma^{-2.2}$ ($\gamma$ is the Lorentz factor
of the cosmic ray), as suggested by observations of synchrotron
emission from shocks \citep{2006ApJ...652.1508S}.  Furthermore,
several strongly magnetized neutron stars appear to have emission
similar to pulsar wind nebulae
\citep{2009ApJ...703L..41R,2003ApJ...582..783H,2003ApJ...591L.143G}.
The apparent absence of such structures around the classical magnetars
(the soft-gamma repeaters and anomalous x-ray pulsars) may simply owe
to the difficulty of detecting them at their relatively larger
distances.  Perhaps these nebulae are powered by pulsar spin-down;
however, the efficiency of the conversion of spin-down energy to
nebular emission appears to be higher than in typical pulsar wind
nebulae, so magnetic field decay may play a role
\citep{2009ApJ...703L..41R}.  Moreover, magnetars may be born with
spin rates characteristic of typical pulsars \citep[or even
faster;][]{Thom93b} so in their youth they may emit pairs in a similar
manner as pulsars (of course, the time evolution of the emission would
differ but this only has a second-order effect on the observed
spectra).  Therefore, with supporting hints and without definite
observational evidence to the contrary, we assume that the magnetars
emit a similar spectrum to pulsars.  This spectrum is softer than that
adopted by \citet{2008arXiv0812.4457P} who used $\propto
\gamma^{-1.4}$ through $\gamma^{-2}$.  The spectrum of particles
produced is taken to extend from $\gamma=1$ to $\gamma=10^9$.

\subsection{Diffusion}

Like \citet{2008arXiv0812.4457P}, we use the model of
\citet{1995PhRvD..52.3265A} to calculate the diffusion of cosmic rays
through the Galaxy.  We will briefly summarize previous results and
the differences between our treatment and that of
\citet{2008arXiv0812.4457P}.

Because of the Galactic magnetic field, cosmic rays cannot travel
directly from their source to Earth.  Rather, they diffuse through the
Galaxy with higher energy particles having larger Larmor radii
diffusing more quickly than lower energy particles.  Meanwhile, as the
particles gyrate about magnetic field lines, they also lose energy.
If the region over which the particles diffuse is larger than the
volume containing the important sources, it is appropriate to assume
that the diffusion is spherically symmetric.  This yields the following
equation for the energy distribution of the cosmic rays
 \citep{1995PhRvD..52.3265A,2008arXiv0812.4457P},
\begin{equation}
 \label{eq:1}
\frac{\partial f}{\partial t} = \frac{D(\gamma)}{r^2}
\frac{\partial}{\partial r} r^2 \frac{\partial f}{\partial r} + 
\frac{\partial \left ( P(\gamma) f \right )}{\partial \gamma} + Q \, ,
\end{equation}
where $D(\gamma)$ is the diffusion coefficient, $P(\gamma)$ is the
energy loss rate of the particles, and $Q$ is the source term.
Ionization, bremsstrahlung, inverse Compton, and synchrotron radiation
draw energy from the cosmic rays as they travel, but for cosmic rays 
younger than $10^7$~yr the final two processes dominate.  In this regime
we have
\begin{equation}
P(\gamma) = p_2 \gamma^2 \simeq 5.2 \times 10^{-21} \gamma^2 {\rm s}^{-1}
 \label{eq:2}
\end{equation}
and a simple solution to Eq.~\ref{eq:1} follows
for an instantaneous burst of cosmic rays at the origin,
\begin{equation}
f(r,t,\gamma) = \frac{N_0 \gamma^{-\alpha}}{\pi^{3/2} r^3} \left ( 1 -
 p_2 t \gamma \right )^{\alpha-2} \left (\frac{r}{r_{\rm dif}}
\right)^3 e^{-\left (r/r_{\rm dif}\right)^2}
 \label{eq:3}
\end{equation}
for $\gamma < \gamma_{\rm cut} = \left (p_2 t\right)^{-1}$ (otherwise
$f$ vanishes).  In this equation $N_0$ is the total number of cosmic
rays produced with a spectrum of $dN/d\gamma \propto
\gamma^{-\alpha}$; here we take $\alpha=2.2$.   The diffusion length
is given approximately by
\begin{equation}
 \label{eq:4}
 r_{\rm dif}(\gamma,t) \simeq 2 \sqrt{ D(\gamma)t
   \frac{1-\left(1-\gamma/\gamma_{\rm cut}\right
     )^{1-\delta}}{\left(1-\delta\right )\gamma/\gamma_{\rm cut}}}. 
\end{equation}
In contrast with  \citet{2008arXiv0812.4457P}, we do not choose the
diffusion coefficient $D(\gamma)$ to be a strict power law but 
assume rather 
that it
saturates below several GeV  \citep{1995PhRvD..52.3265A}
\begin{equation}
 \label{eq:5}
 D(\gamma) = D_0 \left ( 1 + \gamma/\gamma_*\right )^\delta
\end{equation}
where we take $\gamma_* = (3 {\rm GeV})/m_e c^2$.  In addition to
giving the value of $D_0$, Table~\ref{tab:diff} presents the values
of $D(\gamma)$ at 10~GeV as $D_{10}$ and the value at 1~GeV if one
extrapolates the power-law behavior at high energy downward.  The
latter is for comparison with  \citet{2008arXiv0812.4457P}, who uses a
strict power-law relation.
\begin{table}
\caption{Parameters of the diffusion calculations for the two
scenarios.  $D_{10}$ is the value of the diffusion coefficient at
10~GeV, and $D_1$ is value at 1~GeV if one extrapolates 
power-law behavior at high energies to 1~GeV 
 \citep[for comparison with][]{2008arXiv0812.4457P}. Both are in units
 of cm$^2$/s. The pair-production efficiencies for pulsars is
 one-tenth that of magnetars in both cases.}
\label{tab:diff}
\begin{tabular}{lcccccc}
\hline
 & $D_{10}$ & $D_1$ & $D_0$ & $\delta$ & $\epsilon_m$ \\

\hline
LOW & $10^{28}$ & $2.1\times 10^{27}$ & $4.1\times 10^{27}$ & 0.6 &
0.1 \\
HIGH & $7.5\times 10^{28}$ & $2.1\times 10^{28}$ & $3.6\times 10^{28}$
& 0.5 & 0.3 \\
\hline
\end{tabular}
\end{table}

\subsection{Distribution}

We adopt the analysis of \citet{Gill07} to determine the expected
spatial locations of magnetars and pulsars. This distribution has two
components.  One owes to the high mass stars throughout the Galaxy
\citep{2000AJ....120..314R,2005AJ....130.1652R}. This distribution
decays exponentially from the center of the Galaxy with a scale length
of 2.8~kpc and vertically away from the galactic disk with a scale
height of 45~pc. The second component is the overabundance of high
mass stars known as the Gould belt.  Many of the nearby middle-aged
neutron stars are probably associated with this recent star formation
\citep{2003A&A...406..111P}. Including this population is critical
because nearby young sources can dominate the observed spectrum of
cosmic rays.  To determine the ages of the sources, we assume that the
birthrate of pulsars and magnetars has been uniform in time throughout
the Galaxy, so we simply choose 20,000 ages from a uniform
distribution between zero and $10^6$~years for pulsars and
$10^7$~years for magnetars.

Although we examine sources with ages up to $10^7$~years, the bulk of
the emission emerges within the first 10,000 years for magnetars or
shorter for pulsars.  Therefore, even with a kick of up to 2,000~km/s,
the sources move at most twenty parsecs from their birthplaces during
their active phases.  It is reasonable to neglect the motion of the
sources through the Galaxy during their active period.  This contrasts
with \citet{2008arXiv0812.4457P} who used present day positions of
pulsars, which leads to a more diffuse distribution (e.g. more
extended perpendicular to the Galactic disk) than when the sources
were active because of natal kicks given to neutron stars.

\section{Results}

There are two important questions to address.  The first is whether
the magnetic fields of the magnetars carry sufficient energy to
account for the observed fluxes of cosmic rays.  The second is whether
or not magnetars can account for this observed spectrum of cosmic rays
incident upon the Earth, given the assumed spectrum of cosmic ray
production.  We are interested in both the mean cosmic ray abundance
over the realizations but also the range.  Understanding the range is
crucial because one or two nearby young sources can dominate the
abundances, and we may simply be living in a somewhat special place at
a somewhat special time.  It is important to mention that our results
even for pulsars do not agree with those of
\citet{2008arXiv0812.4457P} because of the aforementioned differences
in modeling.
\begin{figure}
\includegraphics[width=8.5cm,clip]{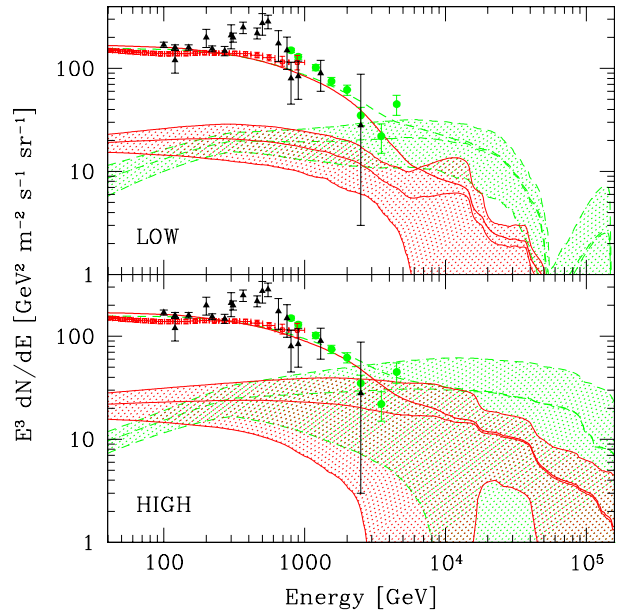}
\caption{Spectrum of cosmic rays from pulsars (green dashed
 curves) and magnetars (red solid curves) within the Galaxy
 under the ``LOW'' and ``HIGH'' diffusion scenarios.  
 For each type of object, the shaded region gives one-standard deviation
 variation over the realisations.
 Furthermore, the bold curve gives the mean plus the results of 
\citet{2010arXiv1002.1910D} excluding the contribution of pulsars.
 The  solid triangles give the results from
 ATIC \citep{2008Natur.456..362C}, the circles give the results
 from HESS \citep{2008PhRvL.101z1104A} and the squares give the
 results from Fermi \citep{2009PhRvL.102r1101A}.  For each scenario the pair
 production efficiency is set so that the total cosmic-ray abundance
 does not exceed that observed by Fermi around 300~GeV. }
\label{fig:all}
\end{figure}

Fig.~\ref{fig:all} shows the expected abundance of cosmic rays for
pulsars and magnetars for the two different diffusion scenarios.
Although the total energy reservoir is the same for a particular
pulsar or magnetar, the magnetars are ten-times less numerous, so they
are assigned a ten times larger efficiency so that the total
cosmic-ray production is the same for the two models.  The efficiency
is set for the two diffusion scenarios by insisting that the model
predictions agree with the Fermi data around 300~GeV (see
Tab.~\ref{tab:diff}).  This also where the pulsar and magnetar models
cross.

For both diffusion scenarios and both populations the resulting cosmic
ray abundance comes from nearby old objects and more distant young
ones.  The Galactic distribution of the sources goes from
approximately three-dimensional to disk-like around a distance of
1~kpc.  Given the abundance of pulsars and magnetars, the youngest
source within this distance is typically $10^4$~yr or $10^5$~yr for
pulsars or magnetars respectively.  This yields cutoff energies of
approximately 31~TeV and 3.1~TeV, roughly where the flux begins to
decrease in the ``LOW'' scenario.  Of course, the cutoff energy does
not depend on the diffusion scenario.  In the ``LOW'' scenario the
diffusion length is about 1~kpc near the cutoff so there are few young
sources to contribute to the flux at Earth at these energies, and the
flux drops and recovers at high energies as the diffusion length
increases.

On the other hand, in the ``HIGH'' scenario the diffusion length is a
few~kpc, so we receive cosmic rays from further in the disk from
younger sources even around the local cutoff --- as the older, nearby
sources cut out, the younger more distant sources seamlessly take
over.  The transition between these two populations leaves an imprint
on the expected spectrum.  The energy of this imprint depends on the
details of the diffusion process and the age of the sources. As the
diffusion becomes less rapid, and the objects become older, the
transition energy decreases. For the magnetar population in the
``LOW'' scenario, this transition occurs tantalizingly close to the
dip in the HESS data.  \citet{2009A&A...508..561A} interpret this dip
as a large steepening in the power-law around 1~TeV consistent with
the magnetar results.  The pulsars do not follow the dip as sharply
but they better fit the final HESS data point at about 5~TeV.
 
\begin{figure}
\includegraphics[width=8.5cm,clip]{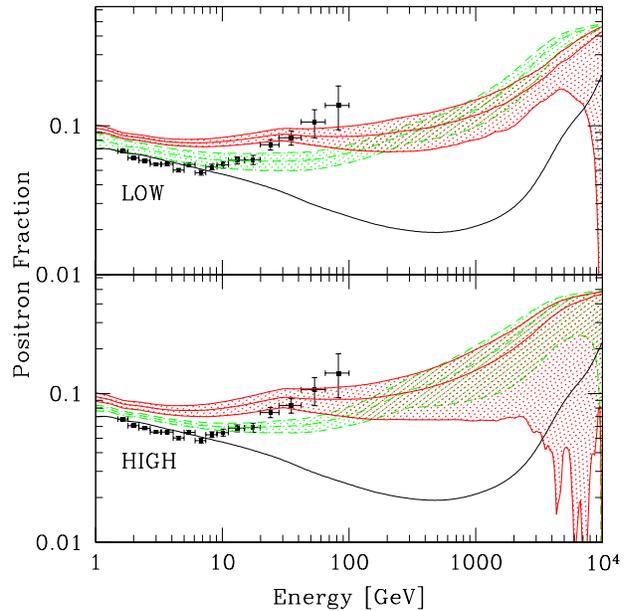}
\caption{The observed positron fraction of cosmic rays from pulsars
  (green dashed curves) or magnetars (red solid curves) combined with
  a standard model for cosmic-ray production and propagation exclusive
  of neutron stars from \citet{2010arXiv1002.1910D} (given by the
  black curves) under the ``LOW'' and ``HIGH'' diffusion scenarios.
  For type of object, the shaded region gives one-standard deviation
  variation over the realisations.  The solid squares give the results
  from PAMELA \citep{2009Natur.458..607A}.}
\label{fig:frac}
\end{figure}
Fig.~\ref{fig:frac} shows the expected anti-matter fraction in
electron-positron cosmic rays for the two different diffusion
scenarios compared and combined with a model that excludes a
contribution from either pulsars or magnetars
\citep{2010arXiv1002.1910D}.  The contribution of pulsars or magnetars
increases the expected positron fraction at all energies.  The effect
of magnetars is stronger at lower energies than pulsars because the
positrons received from magnetars are typically older and less
energetic than those from pulsars.  The variation in the expected
positron fraction over realisations is larger for the magnetars
because the observed cosmic rays are produced by fewer dominant
sources, so it is problematic to access which model does better with a
statistical tests.  At higher energies the pulsars predict a larger
positron fraction than magnetars, although here the fractions agree
within the expected variation.  The disagreement with the observations
below 10~GeV although statistically significant is probably not
important because of solar modulation of the positron and electron
flux at low energies that has not been modelled here.

\section{Conclusions}

We have derived the expected spectrum and composition of cosmic rays
from magnetars under the reasonable assumptions that the decay of the
external magnetic field powers the cosmic-ray production and that the
birthrate of magnetars is about ten percent of that of pulsars.  We
assume that pulsar spin-down powers the cosmic-ray production from
these more weakly magnetized neutron stars.  We find that either the
ensemble of magnetars or that of pulsars can reproduce many of the
puzzling features of the spectrum of electron-positron cosmic rays
recently discovered by ATIC and HESS.  The differences in the
appearance and location of these features in the cosmic-ray abundance
between the models results mainly from the the relative rarity of
magnetars.  Although the relative heights of the two bumps in cosmic
rays detected at Earth do depend on the the assumed power-law
production spectrum, the location of the transition between them does
not. On the other hand, the ensemble of pulsars cannot easily
reproduce the sharp dip in abundance around 2~TeV found by HESS --
pulsars do reproduce the upturn at 5~TeV, so a combination of the two
ensembles may explain both the dip and the upturn.  Both pulsars and
magnetars can account for the positron abundance in cosmic rays with
energies of about one to one-hundred GeV. With additional modelling of
the cosmic-ray diffusion, energy losses and production, even more
detailed agreement with the observations can be achieved.

%\vspace{-0.3in}
\section*{Acknowledgments}
%\begin{acknowledgments}
This research was supported by funding from NSERC, CFI and BCKDF and
made use of the NASA ADS.
%\end{acknowledgments}

\bibliographystyle{mn2e}
\bibliography{math,mine,physics,cosmic,ns}

\begin{thebibliography}{}

\bibitem[\protect\citeauthoryear{{Abdo} et~al.,}{{Abdo}
  et~al.}{2009}]{2009PhRvL.102r1101A}
{Abdo} A.~A.,  et~al., 2009, Physical Review Letters, 102, 181101

\bibitem[\protect\citeauthoryear{{Adriani} et~al.,}{{Adriani}
  et~al.}{2009}]{2009Natur.458..607A}
{Adriani} O.,  et~al., 2009, \nat, 458, 607

\bibitem[\protect\citeauthoryear{{Aharonian} et~al.,}{{Aharonian}
  et~al.}{2008}]{2008PhRvL.101z1104A}
{Aharonian} F.,  et~al., 2008, \prl, 101, 261104

\bibitem[\protect\citeauthoryear{{Aharonian} et~al.,}{{Aharonian}
  et~al.}{2009}]{2009A&A...508..561A}
{Aharonian} F.,  et~al., 2009, \aap, 508, 561

\bibitem[\protect\citeauthoryear{{Atoyan}, {Aharonian} \& {V{\"o}lk}}{{Atoyan}
  et~al.}{1995}]{1995PhRvD..52.3265A}
{Atoyan} A.~M.,  {Aharonian} F.~A.,    {V{\"o}lk} H.~J.,  1995, \prd, 52, 3265

\bibitem[\protect\citeauthoryear{{Beloborodov} \& {Thompson}}{{Beloborodov} \&
  {Thompson}}{2007}]{2007ApJ...657..967B}
{Beloborodov} A.~M.,  {Thompson} C.,  2007, \apj, 657, 967

\bibitem[\protect\citeauthoryear{{Chang}, {Adams}, {Ahn}, {Bashindzhagyan},
  {Christl}, {Ganel}, {Guzik}, {Isbert}, {Kim}, {Kuznetsov}, {Panasyuk},
  {Panov}, {Schmidt}, {Seo}, {Sokolskaya}, {Watts}, {Wefel}, {Wu} \&
  {Zatsepin}}{{Chang} et~al.}{2008}]{2008Natur.456..362C}
{Chang} J.,  {Adams} J.~H.,  {Ahn} H.~S.,  {Bashindzhagyan} G.~L.,  {Christl}
  M.,  {Ganel} O.,  {Guzik} T.~G.,  {Isbert} J.,  {Kim} K.~C.,  {Kuznetsov}
  E.~N.,  {Panasyuk} M.~I.,  {Panov} A.~D.,  {Schmidt} W.~K.~H.,  {Seo} E.~S.,
  {Sokolskaya} N.~V.,  {Watts} J.~W.,  {Wefel} J.~P.,  {Wu} J.,    {Zatsepin}
  V.~I.,  2008, \nat, 456, 362

\bibitem[\protect\citeauthoryear{{Delahaye}, {Lavalle}, {Lineros}, {Donato} \&
  {Fornengo}}{{Delahaye} et~al.}{2010}]{2010arXiv1002.1910D}
{Delahaye} T.,  {Lavalle} J.,  {Lineros} R.,  {Donato} F.,    {Fornengo} N.,
  2010, ArXiv e-prints, 1002.1910

\bibitem[\protect\citeauthoryear{Duncan \& Thompson}{Duncan \&
  Thompson}{1992}]{Dunc92}
Duncan R.~C.,  Thompson C.,  1992, ApJL, 392, 9

\bibitem[\protect\citeauthoryear{Gill \& Heyl}{Gill \& Heyl}{2007}]{Gill07}
Gill R.,  Heyl J.,  2007, \mn, 381, 52

\bibitem[\protect\citeauthoryear{{Gonzalez} \& {Safi-Harb}}{{Gonzalez} \&
  {Safi-Harb}}{2003}]{2003ApJ...591L.143G}
{Gonzalez} M.,  {Safi-Harb} S.,  2003, \apjl, 591, L143

\bibitem[\protect\citeauthoryear{{Helfand}, {Collins} \& {Gotthelf}}{{Helfand}
  et~al.}{2003}]{2003ApJ...582..783H}
{Helfand} D.~J.,  {Collins} B.~F.,    {Gotthelf} E.~V.,  2003, \apj, 582, 783

\bibitem[\protect\citeauthoryear{Heyl \& Hernquist}{Heyl \&
  Hernquist}{1997a}]{Heyl97kes}
Heyl J.~S.,  Hernquist L.,  1997a, \apjl, 489, 67

\bibitem[\protect\citeauthoryear{Heyl \& Hernquist}{Heyl \&
  Hernquist}{1997b}]{Heyl97magnetar}
Heyl J.~S.,  Hernquist L.,  1997b, \apjl, 491, 95

\bibitem[\protect\citeauthoryear{Heyl \& Hernquist}{Heyl \&
  Hernquist}{1998}]{Heyl98rxj}
Heyl J.~S.,  Hernquist L.,  1998, \mn, 297, L69

\bibitem[\protect\citeauthoryear{Heyl \& Hernquist}{Heyl \&
  Hernquist}{2005a}]{Heyl05sgr}
Heyl J.~S.,  Hernquist L.,  2005a, \mn, 362, 777

\bibitem[\protect\citeauthoryear{Heyl \& Hernquist}{Heyl \&
  Hernquist}{2005b}]{Heyl03sgr}
Heyl J.~S.,  Hernquist L.,  2005b, \apj, 618, 463

\bibitem[\protect\citeauthoryear{Heyl \& Kulkarni}{Heyl \&
  Kulkarni}{1998}]{Heyl98decay}
Heyl J.~S.,  Kulkarni S.~R.,  1998, \apjl, 506, 61

\bibitem[\protect\citeauthoryear{{Hooper}}{{Hooper}}{2009}]{2009NatPh...5..176%
H}
{Hooper} D.,  2009, Nature Physics, 5, 176

\bibitem[\protect\citeauthoryear{{Kuiper}, {Hermsen}, {den Hartog} \&
  {Collmar}}{{Kuiper} et~al.}{2006}]{2006ApJ...645..556K}
{Kuiper} L.,  {Hermsen} W.,  {den Hartog} P.~R.,    {Collmar} W.,  2006, \apj,
  645, 556

\bibitem[\protect\citeauthoryear{{Mereghetti}, {G{\"o}tz}, {Mirabel} \&
  {Hurley}}{{Mereghetti} et~al.}{2005}]{2005A&A...433L...9M}
{Mereghetti} S.,  {G{\"o}tz} D.,  {Mirabel} I.~F.,    {Hurley} K.,  2005, \aap,
  433, L9

\bibitem[\protect\citeauthoryear{{Popov}, {Colpi}, {Prokhorov}, {Treves} \&
  {Turolla}}{{Popov} et~al.}{2003}]{2003A&A...406..111P}
{Popov} S.~B.,  {Colpi} M.,  {Prokhorov} M.~E.,  {Treves} A.,    {Turolla} R.,
  2003, \aap, 406, 111

\bibitem[\protect\citeauthoryear{{Profumo}}{{Profumo}}{2008}]{2008arXiv0812.44%
57P}
{Profumo} S.,  2008, ArXiv e-prints, 0812.4457

\bibitem[\protect\citeauthoryear{{Rea}, {McLaughlin}, {Gaensler}, {Slane},
  {Stella}, {Reynolds}, {Burgay}, {Israel}, {Possenti} \& {Chatterjee}}{{Rea}
  et~al.}{2009}]{2009ApJ...703L..41R}
{Rea} N.,  {McLaughlin} M.~A.,  {Gaensler} B.~M.,  {Slane} P.~O.,  {Stella} L.,
   {Reynolds} S.~P.,  {Burgay} M.,  {Israel} G.~L.,  {Possenti} A.,
  {Chatterjee} S.,  2009, \apjl, 703, L41

\bibitem[\protect\citeauthoryear{{Reed}}{{Reed}}{2000}]{2000AJ....120..314R}
{Reed} B.~C.,  2000, \aj, 120, 314

\bibitem[\protect\citeauthoryear{{Reed}}{{Reed}}{2005}]{2005AJ....130.1652R}
{Reed} B.~C.,  2005, \aj, 130, 1652

\bibitem[\protect\citeauthoryear{{Salmonson}, {Fragile} \&
  {Anninos}}{{Salmonson} et~al.}{2006}]{2006ApJ...652.1508S}
{Salmonson} J.~D.,  {Fragile} P.~C.,    {Anninos} P.,  2006, \apj, 652, 1508

\bibitem[\protect\citeauthoryear{Thompson \& Duncan}{Thompson \&
  Duncan}{1993}]{Thom93b}
Thompson C.,  Duncan R.~C.,  1993, ApJ, 408, 194

\bibitem[\protect\citeauthoryear{Thompson \& Duncan}{Thompson \&
  Duncan}{1995}]{Thom95}
Thompson C.,  Duncan R.~C.,  1995, MNRAS, 275, 255

\bibitem[\protect\citeauthoryear{Thompson \& Duncan}{Thompson \&
  Duncan}{1996}]{Thom96}
Thompson C.,  Duncan R.~C.,  1996, ApJ, 473, 322

\end{thebibliography}
\label{lastpage}
\end{document}